\documentclass[a4paper,UKenglish]{lipics-v2019}

\usepackage{tikz}
\usepackage{booktabs}
\tikzset{big node/.style={circle,fill=white,draw,inner sep=1pt,minimum size=0.6cm}}

\title{Engineering Exact Quasi-Threshold Editing}

\author{Lars Gottesbüren}{Institute of Theoretical Informatics, Karlsruhe Institute of Technology, Karlsruhe, Germany}{lars.gottesbueren@kit.edu}{https://orcid.org/0000-0003-1895-5828}{}
\author{Michael Hamann}{Institute of Theoretical Informatics, Karlsruhe Institute of Technology, Karlsruhe, Germany}{michael.hamann@kit.edu}{https://orcid.org/0000-0002-6958-4927}{}
\author{Philipp Schoch}{Institute of Theoretical Informatics, Karlsruhe Institute of Technology, Karlsruhe, Germany}{}{}{}
\author{Ben Strasser}{Institute of Theoretical Informatics, Karlsruhe Institute of Technology, Karlsruhe, Germany}{academia@ben-strasser.net}{https://orcid.org/0000-0002-3391-6221}{}
\author{Dorothea Wagner}{Institute of Theoretical Informatics, Karlsruhe Institute of Technology, Karlsruhe, Germany}{dorothea.wagner@kit.edu}{https://orcid.org/0000-0002-9141-7076}{}
\author{Sven Z{\"u}hlsdorf}{Institute of Theoretical Informatics, Karlsruhe Institute of Technology, Karlsruhe, Germany}{zuehlsdorf.kit@ghostdub.de}{}{}

\authorrunning{L. Gottesbüren, M. Hamann, P. Schoch, B. Strasser, D. Wagner and S. Z{\"u}hlsdorf}
\Copyright{Lars Gottesbüren, Michael Hamann, Philipp Schoch, Ben Strasser, Dorothea Wagner and Sven~Z{\"u}hlsdorf}



\supplement{Implementation: \url{https://github.com/kit-algo/fpt-editing}}

\funding{This work was supported by the Deutsche Forschungsgemeinschaft (DFG, German Research Foundation) under grants WA654/19-2 and WA654/22-2. The authors acknowledge support by the state of Baden-Württemberg through bwHPC.}

\acknowledgements{We thank James Nastos and Mark Ortmann for helpful discussions.}
\nolinenumbers 
\hideLIPIcs  
\EventEditors{Simone Faro and Domenico Cantone}
\EventNoEds{2}
\EventLongTitle{18th International Symposium on Experimental Algorithms (SEA 2020)}
\EventShortTitle{SEA 2020}
\EventAcronym{SEA}
\EventYear{2020}
\EventDate{June 16--18, 2020}
\EventLocation{Catania, Italy}
\EventLogo{}
\SeriesVolume{160}
\ArticleNo{7}

\begin{document}

\maketitle

\begin{abstract}
  Quasi-threshold graphs are $\{C_4, P_4\}$-free graphs, i.e., they do not contain any cycle or path of four nodes as an induced subgraph.
  We study the $\{C_4, P_4\}$-free editing problem, which is the problem of finding a minimum number of edge insertions or deletions to transform an input graph into a quasi-threshold graph.
  This problem is NP-hard but fixed-parameter tractable (FPT) in the number of edits by using a branch-and-bound algorithm and admits a simple integer linear programming formulation (ILP).
  Both methods are also applicable to the general $\mathcal{F}$-free editing problem for any finite set of graphs~$\mathcal{F}$.
  For the FPT algorithm, we introduce a fast heuristic for computing high-quality lower bounds and an improved branching strategy.
  For the ILP, we engineer several variants of row generation.
  We evaluate both methods for quasi-threshold editing on a large set of protein similarity graphs.
  For most instances, our optimizations speed up the FPT algorithm by one to three orders of magnitude.
  The running time of the ILP, that we solve using Gurobi, becomes only slightly faster.
  With all optimizations, the FPT algorithm is slightly faster than the ILP, even when listing all solutions.
  Additionally, we show that for almost all graphs, solutions of the previously proposed quasi-threshold editing heuristic QTM are close to optimal.
 \end{abstract}

\section{Introduction}

We study graph edge editing problems.
The \emph{distance} between two graphs $G$ and $H$, with the same node set, is the minimum number of edge insertions or deletions needed to transform $G$ into $H$.
Given a graph class $\mathcal{C}$ and a graph $G$, the editing problem asks for a graph $H\in \mathcal{C}$ closest to $G$.
The corresponding decision problem asks whether $k$ edits are sufficient to transform $G$ into a graph $H \in \mathcal{C}$.
We study algorithms to solve editing problems exactly.

A graph $H$ is an \emph{induced subgraph} of a graph $G$, if there exists an injective mapping $\pi$ from the nodes of $H$ onto the nodes of $G$ such that there is an edge between two nodes of $H$, if and only if there is an edge between the corresponding nodes in $G$.
A graph $G$ that does not contain $H$ as induced subgraph is \emph{$H$-free}.
Analogously, for a set of forbidden subgraphs $\mathcal{F}$, a graph $G$ is $\mathcal{F}$-free, if no graph $H\in\mathcal{F}$ is an induced subgraph of $G$.

\begin{figure}[btp]
  \begin{subfigure}[b]{.16\linewidth}
    \centering
    \begin{tikzpicture}[every node/.style={big node}]
      \draw (5,-0.5) node {} -- (6, -0.5) node {} -- (6, 0.5) node {} -- (5, 0.5) node {} -- (5, -0.5);
    \end{tikzpicture}
    \caption{A $C_4$.}\label{fig:c4}
  \end{subfigure}
  \begin{subfigure}[b]{.16\linewidth}
    \centering
    \begin{tikzpicture}[every node/.style={big node}]
      \draw (0,1) node {} -- (0, 0) node {} -- (1, 0) node {} -- (1, 1) node {};
    \end{tikzpicture}
    \caption{A $P_4$.}\label{fig:p4}
  \end{subfigure}
  \begin{subfigure}[b]{.38\linewidth}
    \centering
    \begin{tikzpicture}[every node/.style={big node}]
      \draw (1, 0) node (bottom left) {$b$} -- (2,0.5) node (center) {$c$} -- (3, 0) node (bottom right) {$d$};
      \draw (bottom left) -- (1, 1) node (top left) {$a$} -- (center) -- (3, 1) node (top right) {$e$} -- (bottom right);
      \draw (bottom left) -- (bottom right);
      \draw (top left) -- (top right);
    \end{tikzpicture}
    \caption{A graph with node-induced $C_4$.}\label{fig:graph_with_c4}
  \end{subfigure}
  \begin{subfigure}[b]{.26\linewidth}
    \centering
    \begin{tikzpicture}[every node/.style={big node}]
      \draw (0, 0) node (left) {$d$} -- (1, 0) node (center) {$e$} -- (2, 0) node (right) {$f$};
      \draw (left) -- (0, 1) node (top left) {$a$} -- (1, 1) node (top center) {$b$} -- (center);
      \draw (top left) -- (center);
      \draw (top center) -- (left);
      \draw (center) -- (2, 1) node (top right) {$c$};
    \end{tikzpicture}
    \caption{A $\{C_4, P_4\}$-free graph.}\label{fig:p4c4_free_graph}
  \end{subfigure}
  \caption{A $C_4$, a $P_4$ and examples for graphs that are (not) $\{C_4,P_4\}$-free.}
\end{figure}
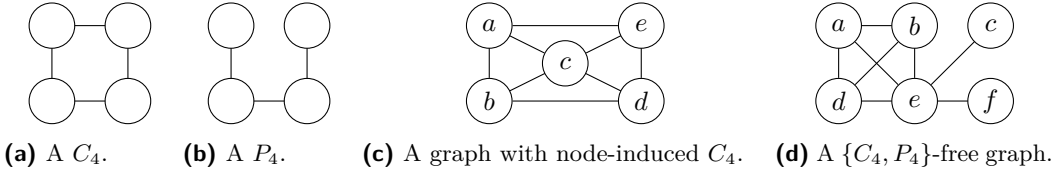

We denote by $P_\ell$ a path graph with $\ell$ nodes.
Similarly, $C_\ell$ denotes a cycle graph with $\ell$ nodes.
Figures~\ref{fig:c4} and~\ref{fig:p4} depict a $C_4$ and a $P_4$.
The graph depicted in Figure~\ref{fig:graph_with_c4} is not $\{C_4,P_4\}$-free as the nodes $(a, b, d, e)$ form an induced $C_4$.
In contrast, the graph depicted in Figure~\ref{fig:p4c4_free_graph} is $\{C_4,P_4\}$-free.
The nodes $(a, b, e, c)$ form a $P_4$, however, as there is an edge between $a$ and $e$, the subgraph is not an induced subgraph.

While the theoretical part of our study considers any $\mathcal{F}$-free edge editing problem for a finite set of subgraphs $\mathcal{F}$, our experimental study considers $\{C_4,P_4\}$-free graphs.
These are also called \emph{quasi-threshold} or \emph{trivially perfect} graphs.
Quasi-threshold editing has applications in detecting communities in social friendship networks.
Nastos and Gao~\cite{ng-f-13} detect communities in a graph $G$ by computing a closest quasi-threshold graph $H$ of $G$.
Each connected component in $H$ corresponds to a community in $G$.

For many choices of $\mathcal{F}$, $\mathcal{F}$-free editing is NP-hard, in particular for $\mathcal{F} = \{C_4,P_4\}$~\cite{ng-f-13}.
The $\mathcal{F}$-free edge editing problem is fixed-parameter tractable (FPT) in the number of edits $k$~\cite{c-f-96}.
This proof directly leads to a branch-and-bound algorithm, see Section~\ref{sec:base_fpt}.
For $\{C_4, P_4\}$-free edge editing, it has a running time of $O(6^k \cdot (n + m))$, where $n$ and $m$ are the number of nodes and edges.
Unfortunately, social networks typically require a large number of edits~\cite{bhsw-fqte-15}, which makes plain FPT algorithms impracticable.
Therefore, in~\cite{ng-f-13} and~\cite{bhsw-fqte-15}, quasi-threshold editing heuristics have been introduced for detecting communities in social friendship networks.
However, as both approaches are heuristics, they might detect communities that are different from those defined by the model that assumes an optimal solution.
Our goal is to improve the running time of exact $\{C_4,P_4\}$-free editing in practice in order to make it feasible at least for small networks.
This allows us to study exact solutions of the community detection problem and to verify the quality of heuristics.

\subsection{Related Work}

For the special case of $\{C_4,P_4\}$-free edge deletion, where only edge deletion operations are allowed, optimized branching rules have been proposed that reduce the running time of the trivial algorithm from $O(4^k \cdot (n + m))$ to $O(2.42^k \cdot (n + m))$~\cite{lwycc-eb-15}.
To the best of our knowledge, for $\{C_4,P_4\}$-free editing, no improved branching rules have been proposed so far.
A polynomial kernel of size $O(k^7)$ for $\{C_4,P_4\}$-free graphs has been proposed~\cite{dp-apktp-17}, which is too large for most practical applications.

A frequently considered problem is $\{P_3\}$-free editing, better known as cluster editing~\cite{bb-ce-13}.
Early approaches for cluster editing include a linear programming formulation with cutting planes that are incrementally added (in batches of a few hundred constraints)~\cite{gw-a-89}.
Later, exact algorithms based on integer linear programming as well as kernelization and more efficient FPT algorithms have been considered~\cite{bbk-eacee-11}.
In~\cite{hh-pompa-15}, the authors combine the FPT algorithm with kernelization as well as upper and lower bounds.
Editing to $\{P_4\}$-free graphs has been considered in phylogenomics~\cite{hwllms-p-15} using a simple ILP-based approach.

In a bachelor thesis~\cite{b-gzp-15}, $\{P_5\}$-free editing has been considered for community detection.
They apply lower bounds, data reduction rules and rules for disallowing certain edits.

\subsection{Our Contribution}

In this paper, we compare two different methods for solving $\mathcal{F}$-free editing problems.
The first is a branch-and-bound FPT algorithm while the second is an ILP.
For the FPT algorithm, we propose a novel lower bound algorithm based on local search heuristics for independent sets as well as an improved branching strategy.
Additionally, we parallelize our implementation.
For the ILP, we engineer several variants of row generation.
We assess the running time improvements of the different optimizations for quasi-threshold editing on a large benchmark set of 716 graphs that are connected components of a protein similarity graph.
This benchmark set has previously been used to evaluate cluster editing algorithms~\cite{rwbmtb-ehawc-07,bbbt-awce-08}.
On 75\% of the instances, our improved bounds and optimized branching choices yield speedups of one to three orders of magnitude for the FPT algorithm.
For the ILP, we are only able to achieve small speedups.
With all optimizations, in the median, the FPT algorithm is twice as fast as the ILP, even when enumerating all possible optimal solutions exactly once.
Compared with the parallel execution of Gurobi~\cite{gurobi}, the FPT algorithm achieves better speedups.
Additionally, we evaluate an LP relaxation as lower bound.
We prove that its bounds are at least as good as our local search bounds.
In our experiments, however, it is too slow to be competitive.

Further, we compare our exact solutions with heuristic solutions found by QTM~\cite{bhsw-fqte-15}.
It turns out that many heuristic solutions are exact and all but one of them are close to the exact solution.
Additionally, we are able to solve four out of the five social networks considered in~\cite{ng-f-13}, of which only one was solved previously~\cite{n-u-15}.

\subsection{Outline}

We start by introducing the preliminaries in Section~\ref{sec:preliminaries}.
We describe the ILP formulation and the optimizations we apply to it in Section~\ref{sec:ilp}.
In Section~\ref{sec:base_fpt}, we then introduce the branch-and-bound FPT algorithm including existing and novel optimizations.
In Section~\ref{sec:experiments}, we present our experimental setup and evaluation.
We conclude in Section~\ref{sec:conclusion}.

\section{Preliminaries}\label{sec:preliminaries}

All graphs in this paper are undirected, unweighted, and finite.
Further, no graph has self-loops or multi-edges.
A graph $G = (V_G, E_G)$ consists of $n := |V_G|$ nodes and $m := |E_G|$ undirected edges.
By $\overline{E_G}$, we denote the complement of the edges.
In the following, $k$ denotes the maximum number of edits.

\section{Integer Linear Programming}\label{sec:ilp}

In this section, we describe an ILP formulation for $\mathcal{F}$-free editing that is based on an existing formulation for cluster editing~\cite{gw-a-89}.
Further, we introduce our optimizations based on row generation and modified constraints to make the ILP practical for small instances.

For every node pair $u,v \in {V_G \choose 2}$ we introduce a variable $x_{uv} \in \{0,1\}$ which is $1$ if the node pair is an edge in the edited graph and $0$ otherwise.
We add constraints to ensure that no forbidden subgraph $H \in \mathcal{F}$ can be induced in $G$ via an injective node mapping $\pi$:

\begin{equation}\label{ilp:destroy_subgraph}
  \forall H \in \mathcal{F}, \forall \pi \colon V_H \hookrightarrow V_G: \sum_{\{u, v\} \in E_H} (1 - x_{\pi(u) \pi(v)}) + \sum_{\{u, v\} \in \overline{E_H}} x_{\pi(u) \pi(v)} \geq 1
\end{equation}

The objective minimizes the number of edits:
\begin{equation}\label{ilp:objective}
\min \sum_{\{u,v\} \in E_G} (1 - x_{uv}) + \sum_{\{u, v\} \in \overline{E_G}} x_{uv}
\end{equation}

\subsection{Row Generation}\label{sec:ilp:rowgeneration}

Generating all of the above-mentioned constraints is infeasible, even for small instances.
Row generation (also called lazy constraints) aims to speed up ILP solvers by starting with a small subset of the constraints and subsequently adding constraints that are violated in intermediate solutions.
We start with constraints for forbidden subgraphs in the input graph.
In our experiments, we consider two options to add constraints violated in an intermediate solution: adding either all violated constraints or only one.

The ILP solver uses LP relaxations to prune its search.
These can be strengthened by adding constraints from Equation~\ref{ilp:destroy_subgraph} that are violated by the LP relaxation.
We generate constraints in three steps.
First, we consider each node pair $\{u, v\}$ for which the relaxation solution has a value different from the input graph.
We edit it, then enumerate the forbidden subgraph embeddings containing $u$ and $v$, add the constraint that is most violated (i.e., whose left side is furthest below $1$) and then revert the edit.
Ties are broken uniformly at random.
Second, we apply the same procedure to the best heuristic solution found so far.
Third, we round the LP solution, i.e., an edge exists iff the corresponding variable is greater than $0.5$.
We then list forbidden subgraph embeddings in this rounded solution and add the corresponding most violated constraint if there is any.
The listing skips forbidden subgraphs for which the corresponding constraint has already been added.

\subsection{Optimizing Constraints for $\{C_4,P_4\}$-free Editing}\label{sec:ilp:specialc4}
If one forbidden subgraph can be transformed into another by a single edit, we can omit a node pair from the constraint for this subgraph.
This is similar to the optimization described in Section~\ref{sec:cons-edit-last}.
For a $P_4$, this is the node pair consisting of the two degree-one nodes.
For a $C_4$, we can omit any one of its four edges.
We always consider all four possibilities, and in the initial constraint generation as well as the basic row generation variant we add all of them.
With this optimization, the constraints for $C_4$s and $P_4$s are identical.

We can also formulate a constraint for a $C_4$ that explicitly models that two deletions or one insertion are required:

\begin{equation}\label{ilp:c4_constraint}
\forall (u_1, u_2, u_3, u_4) \in V_G^4 : \  0.5 \cdot \sum_{i=1}^4 (1-x_{u_i u_{i+1}}) + x_{u_1 u_3} + x_{u_2 u_4} \geq 1
\end{equation}

\section{The FPT Branch-and-Bound Algorithm}\label{sec:base_fpt}

The FPT algorithm~\cite{c-f-96} is a branch-and-bound algorithm.
For a given maximum number of edits $k$, it either reports that no solution exists or returns a set of $k$ edits.
It works as follows: Find a forbidden subgraph $H$ and branch on all possible edits in $H$.
As $H$ is induced, only edits in $H$ can destroy it and thus one of these edits must be part of the solution.
The algorithm is then recursively called for each branch with $k-1$ remaining edits.

Denote by $p$ the maximum number of nodes in a forbidden subgraph.
Finding $H$ can be done trivially in time $O(n^p)$ by enumerating all subgraphs of the required size.
For specific sets of forbidden subgraphs, such as $\{C_4, P_4\}$, this can be improved to $O(n + m)$~\cite{c-albfs-08,bhsw-fqte-15}.

Every pair of nodes in $H$ is a valid edit.
The branching factor is therefore $p\cdot(p-1)/2$.
The depth of the recursion is bounded by the maximum number of edits $k$.
The total running time is therefore in $O(p^{2k} \cdot n^p)$ for general families of forbidden subgraphs.
For quasi-threshold editing the running time is $O(6^k \cdot (n + m))$.
This can be improved to $O(5^k \cdot (n + m))$ by applying the optimization described in Section~\ref{sec:cons-edit-last}.

For finding the minimum number of edits $k_{\text{opt}}$, the algorithm needs to be executed for increasing values of $k$ until a solution is returned.
For a branching factor of 2 or larger, the running time of all $k < k_{\text{opt}}$ together is at most the running time for $k_{\text{opt}}$.
Thus the total running time is dominated by the running time for $k_{\text{opt}}$.

In the following, we describe several optimizations to reduce the number of explored branches in practice.
We describe existing techniques for avoiding redundant exploration of branches (Section~\ref{sec:avoiding-redundancy}), for skipping certain branches (Section~\ref{sec:cons-edit-last}) as well as lower bounds (Section~\ref{sec:bounds}).
We introduce a novel local search lower bound (Section~\ref{sec:localsearch}), optimized branching choices (Section~\ref{sec:most}), early pruning of branches (Section~\ref{sec:prune}) and a simple parallelization (Section~\ref{sec:parallel}).
In Appendix~\ref{app:implementation-details} we provide in-depth implementation details.

\subsection{Avoiding Redundancy}\label{sec:avoiding-redundancy}

Damaschke~\cite{d-fpece-08} proposes to block node pairs to list every solution exactly once.
When spawning a search tree node $x$ through editing a node pair, it is neither useful to undo that edit in the sub-search-tree rooted at $x$, nor is it useful to perform the edit in sibling search trees.
While this has been introduced for cluster editing, the technique can be applied to arbitrary $\mathcal{F}$-free editing problems.
In Appendix~\ref{app:redundancy}, we explain this technique in detail.

\subsection{Skip Forbidden Subgraph Conversion.}\label{sec:cons-edit-last}

\begin{lemma}
If each forbidden subgraph $A \in \mathcal{F}$ can be transformed into another one $B \in \mathcal{F}$ by one edit, the branching factor of the FPT algorithm can be reduced from ${p \choose 2}$ to ${p \choose 2} - 1$.
\end{lemma}

There is an edit that transforms a $P_4$ into a $C_4$.
Clearly, this edit can be skipped.
Further, there are four edge deletions that transform a $C_4$ into a $P_4$.
One of these can be skipped~\cite{ng-anbsp-10}.
We can choose which one, but as any pair of two edge deletions eliminates the forbidden subgraph, skipping more than one of them might eliminate a necessary branch. 
Since the branching factor is reduced, this decreases the worst-case running time from $O(6^k \cdot (n + m))$ to $O(5^k \cdot (n + m))$ for quasi-threshold editing.

\subsection{Existing Lower Bound Approaches}\label{sec:bounds}

At each branching node, we have a certain number $k$ of edits left.
If we can show that the graph needs at least $k + 1$ edits, we do not need to explore further branches below that node.
Lower bounds have been used for cluster editing~\cite{bbk-eacee-11,hh-pompa-15} and $\{P_5\}$-free editing~\cite{b-gzp-15}.
Commonly, they are based on an LP relaxation of the ILP~\cite{hh-pompa-15}, or on a disjoint packing argument~\cite{b-gzp-15,hh-pompa-15}.

\subparagraph*{Subgraph Packing.}
A \emph{node-pair disjoint subgraph packing} $P$ is a set of induced forbidden subgraphs that do not share a node pair.
As no edit can eliminate more than one subgraph, $|P|$ is a lower bound on the number of edits required.
Taking the previously mentioned optimizations into account, we can include more subgraphs in $P$ by allowing to share blocked node pairs, as they cannot be edited.
Further, for each forbidden subgraph a node pair that transforms it into another forbidden subgraph may be shared.
In the case of $\mathcal{F} = \{C_4, P_4\}$, the pair of degree-1 nodes of an induced $P_4$ can be shared.
For $C_4$, we can choose any edge to share, but it remains the same as long as the $C_4$ is in the packing.

Finding such a packing can be modeled as an independent set problem~\cite{hh-pompa-15}.
The forbidden subgraphs are nodes and every pair of forbidden subgraphs that shares a non-shareable node pair is connected by an edge.
A natural greedy heuristic for independent sets is to iteratively add the node that has the smallest degree and then remove all its neighbors from the graph.
This can be implemented in linear time by splitting nodes into buckets according to their degree (see e.g.~\cite{arw-flsmisp-12}).
This heuristic has also been used to calculate lower bounds for cluster editing~\cite{hh-pompa-15}.
We are not aware of complexity results of the independent set problem on this special graph class.

In our experiments, we evaluate three bounds based on subgraph packing:
1) A \emph{basic} bound that iteratively adds subgraphs to the packing as they are found.
2) An incremental version of 1) that \emph{updates} the packing as the graph is modified in the branch-and-bound algorithm.
After applying an edit, we remove the subgraph that contains the edited node pair.
After both editing and blocking, we enumerate and add subgraphs to the bound until it is maximal.
3) A greedy bound based on the \emph{minimum degree} heuristic.
In contrast to the first two, this requires storing all forbidden subgraphs.
To avoid this in trivial cases, we first apply 2) to the previous bound and only compute a new packing if this fails to prune the branch.

\subparagraph*{LP relaxation.}
The optimal solution of the LP relaxation provided in Section~\ref{sec:ilp} is an upper bound for the node-pair-disjoint packing problem.
This can be shown by considering an LP with just the constraints that correspond to the subgraphs in a packing.
Each subgraph in the packing is a node-induced subgraph of $G$.
Therefore, the terms on the left side of its corresponding constraint appear in the objective function exactly as they appear in the constraint, confer Equations~\ref{ilp:destroy_subgraph} and~\ref{ilp:objective}.
Each term in the objective function is at least $0$, and each group of terms corresponding to a fulfilled constraint sums to at least $1$.
Since the packing is node-pair disjoint, the constraints do not share any variables and thus groups do not overlap.
Therefore, the objective value is at least the number of subgraphs in the packing.
Adding more constraints can only increase the objective and thus improve the bound.
We can also model blocked node pairs by replacing the corresponding variable by its value.
The variables in the constraints are then disjoint again and thus the same argument applies.

\subsection{Local Search Lower Bound}\label{sec:localsearch}

We propose a lower bound based on a subgraph packing that is computed using an adaptation of the 2-improvements local search heuristic~\cite{arw-flsmisp-12} for independent sets.
Our local search starts with an initial packing and works in rounds.
In each round, it iterates over all forbidden subgraphs in the packing and tries to replace one by two forbidden subgraphs.
If this is not possible, it tries to replace one by one.
Preliminary experiments have shown that choosing this replacement from those candidates which cover the fewest other forbidden subgraphs leads to significantly higher bounds than considering all candidates.
We also found that using this strategy only 70\% of the time and otherwise choosing a random replacement is even better.
We repeat this procedure until in five consecutive rounds only one-by-one replacements were found.
We also terminate the search if the packing remains completely unchanged in a round, or if the packing is large enough to prune the current branch in the search tree.
To make this efficient, we approximate the number of forbidden subgraphs that are covered by a certain forbidden subgraph $H$, by adding up the number of forbidden subgraphs each node pair of $H$ is part of.
For the latter we can efficiently maintain counters.

The initial packing is computed with the basic greedy bound.
For recursive calls, we update the previous bound as discussed above, before employing local search.

\subsection{Branch on Most Useful Node Pairs}\label{sec:most}

We can choose any forbidden subgraph for branching on its possible edits, e.g., the first we find.
If there is a forbidden subgraph with only one non-blocked node pair, we choose it, as this will lead to just one recursive call.
Otherwise, the first node pair we try to edit should ideally lead to a solution, or blocking the edit should prune the search.
We propose to prefer forbidden subgraphs whose non-blocked node pairs are part of many other forbidden subgraphs.
Then, a single edit can eliminate many forbidden subgraphs (possibly leading to a solution) and blocking the node pairs allows adding many subgraphs to the lower bound.
For each forbidden subgraph, we sort its non-blocked node pairs in decreasing order by the number of forbidden subgraphs that contain the respective node pair.
The edits of the selected forbidden subgraph are also tried in this order.
We select the subgraph to branch on using a lexicographical ordering on these counts.
The last node pair is excluded, as there are no branches left to prune.
Additionally, if two subgraphs have identical count sequences (up to the length of the shorter one), we prefer the subgraph with the shorter sequence.

\subsection{Prune Branches Early}\label{sec:prune}

Normally, we attempt to prune a branch after applying an edit and descending into recursion.
With the optimization from Section~\ref{sec:avoiding-redundancy}, the edited node pair of a recursive call remains blocked after returning from recursion.
We update the lower bound to consider this blocked node pair.
If the new lower bound already exceeds the remaining number of edits, we can directly prune all subsequent recursive calls, instead of pruning them individually.
There are two cases for which we skip the bound update to save running time:
If there is only one subsequent recursive call, as we would only prune a single branch, and if the blocked node pair is only part of a single forbidden subgraph, as it cannot yield a better lower bound.

\subsection{Parallelization}\label{sec:parallel}

The algorithm can be parallelized by letting different cores explore different branches.
Due to our optimizations, not every branch needs the same running time.
Therefore, just executing the first branches in parallel is not scalable.
Instead, we use a simple work stealing algorithm.
Whenever a thread has fully explored its branch, it steals a branch on the highest available level from another thread and further explores it.

\section{Experimental Evaluation}\label{sec:experiments}

\newcommand{\fpt}{\textsf{FPT}}
\newcommand{\all}{\textsf{-All}}
\newcommand{\greedy}{\textsf{-G}}
\newcommand{\updatedgreedy}{\textsf{-U}}
\newcommand{\mindegree}{\textsf{-MD}}
\newcommand{\localsearch}{\textsf{-LS}}
\newcommand{\lprelaxation}{\textsf{-LP}}
\newcommand{\bfirst}{\textsf{-F}}
\newcommand{\bmost}{\textsf{-M}}
\newcommand{\bmostpruned}{\textsf{-MP}}

\newcommand{\ilp}{\textsf{ILP}}
\newcommand{\single}{\textsf{-S}}
\newcommand{\basic}{\textsf{-B}}
\newcommand{\relaxation}{\textsf{-R}}
\newcommand{\cfour}{\textsf{-C4}}

\newcommand{\mt}{\textsf{-MT}}

In Appendix~\ref{app:implementation-details} we discuss implementation details.
The C++ source code\footnote{\url{https://github.com/kit-algo/fpt-editing}} of all discussed variants is available online.
We use the C++ interface of Gurobi~\cite{gurobi} to solve ILPs and LPs.
We evaluate our algorithms on a set of 3964 graphs that are connected components of the COG protein similarity data\footnote{\url{https://bio.informatik.uni-jena.de/data/\#cluster_editing_data}} that has already been used for the evaluation of cluster editing algorithms~\cite{rwbmtb-ehawc-07,bbbt-awce-08}.
The dataset consists of a similarity matrix for each graph.
We treat all non-negative scores as edges.
Unless stated otherwise, we restrict our evaluation to the 716 graphs that require at least 20 edits.
On the 3248 excluded graphs, the maximum running time is less than 0.43 seconds for the FPT algorithm using our local search lower bound.
Of these graphs, 1666 require no edits at all.
Further, we evaluate our algorithms on a set of 5 small social networks that were already considered by Nastos and Gao~\cite{ng-f-13}, namely \texttt{karate}~\cite{z-ifmcf-77}, \texttt{grass\_web}~\cite{dhc-spcgf-95}, \texttt{lesmis}~\cite{k-tsgb-93}, \texttt{dolphins}~\cite{lsbhsd-tbdcd-04}, and \texttt{football}~\cite{gn-c-02}.

All experiments were performed on systems with two 8-core Intel Xeon E5-2670 (Sandy Bridge) processors and 64~GB RAM.
We set a global time limit of 1000 seconds.
Experiments comparing just FPT variants were executed on 16 different node orders, running 16 node orders in parallel.
Due to the memory requirements of Gurobi, this is not feasible for the ILP and the LP bound.
For these variants, we run just one instance at a time.
For experiments involving ILP variants, we also limit the experiments to 4 node orders, and, for better  comparability, we run one instance at a time also for the FPT comparison runs in Figure~\ref{fig:bio_final_comparison}.
By default, all algorithms terminate at the first found solution, as the ILP is unable to enumerate solutions.
Variants with the suffix \all~enumerate all solutions.
Further, variants with the suffix \mt{} are parallelized using 16~cores.

\subsection{Variants of the FPT Algorithm}\label{sec:fpt-evaluation}

The baseline branching strategy \bfirst{} uses the first found forbidden subgraph.
Our \texttt{Most} branching strategy from Section~\ref{sec:most} is denoted by \bmost{}, additional early pruning by \bmostpruned{}.
The basic greedy bound is denoted by \greedy{}, the incremental update bound by \updatedgreedy{}, the min-degree heuristic by \mindegree{}, our local search lower bound by \localsearch, and LP relaxations by~\lprelaxation.
The comparison includes the nine variants \fpt\greedy\bfirst\all, \fpt\greedy\bmostpruned\all, \fpt\updatedgreedy\bmostpruned\all, \fpt\mindegree\bfirst\all, \fpt\mindegree\bmostpruned\all, \fpt\lprelaxation\bmostpruned\all, \fpt\localsearch\bfirst\all, \fpt\localsearch\bmost\all{} and \fpt\localsearch\bmostpruned\all.

\begin{figure}[tb]
  \resizebox{\linewidth}{!}{\input{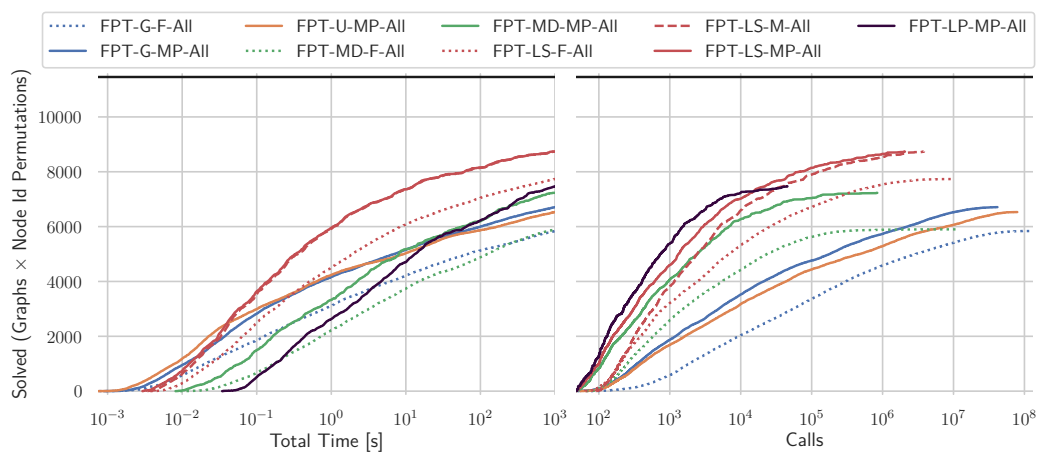}}
  \caption{Number of permutations of graphs of the COG dataset that require at least 20 edits and can be solved within a certain total running time / with a certain number of recursive calls (and extra lower bound updates for \bmostpruned). The horizontal black line indicates the total number of graphs and node permutations that require 20 or more edits, including unsolved instances.
  }\label{fig:cog_times}
\end{figure}

Figure~\ref{fig:cog_times} shows how many of the COG dataset instances can be solved within a certain time limit and with a certain number of recursive calls -- added over all $k$'s.
Additional lower bound calls due to \bmostpruned{} count extra.
An instance is a single node id permutation of a graph, i.e., every graph is counted 16 times.
Of the 716 graphs we are able to solve 547 within the 1000 second time limit.
Below, we also compare calls and running times per instances.

For comparing branching strategies, we fix the local search algorithm \localsearch~as the lower bound.
The median factor of additional calls needed by \bmost{} over \bmostpruned{} is 1.9 and by \bfirst{} over \bmostpruned{} is 3.36, restricted to instances solved by both algorithms.
While the median speedup of \bmostpruned{} over \bfirst{} is 3.11, it is just 1.06 over \bmost{}.
On 5\% of the instances, the speedup is at least 56.62 and 1.24, respectively.
This shows that for \bmost{} the improvement in the number of calls directly leads to similar running time improvements, while early pruning just reduces calls.

For comparing lower bound algorithms, we fix \bmostpruned~as the branching strategy.
There is an inherent trade-off between the number of recursive calls and the time spent per call, with a sweet spot that gives the best overall running time.
The basic greedy bounds need 10 to 24 times as many calls as the other bounds in the median.
The recomputed greedy bound \greedy{} is slightly better than the updated one \updatedgreedy{}, \lprelaxation{} is the best, followed by \localsearch{} and \mindegree{}.

Nonetheless, for very small time limits, \updatedgreedy{} solves the highest number of instances.
For larger time limits, reducing the number of calls pays off, though not at any cost.
The median speedup of min-degree over the LP is 2.16 while needing 47\% more calls in the median.
Local search avoids their substantial memory overhead and spends significantly less time per call than both.
It needs 83\% of the calls of \mindegree{} while being a factor of 12.36 faster in the median.
It is never slower, and on 5\% of the instances even more than 137 times faster than \mindegree{}.

Comparing the state-of-the-art \fpt\mindegree\bfirst\all{} algorithm to our \fpt\localsearch\bmostpruned\all{} algorithm, we need 4.33 times less calls and are 46.06 times faster in the median.
We are never slower, on 75\% of the instances more than 16 times and on 5\% of the instances more than 1044 times faster.
In conclusion, our local search lower bound gives high-quality bounds while being fast.
Our branching rules reduce the running time by another small factor while early pruning mainly reduces the number of calls.
Overall, we achieve a speedup of one to three orders of magnitude over the state-of-the-art.

\subsection{Parallelization}\label{sec:exp:parallelization}

\begin{figure}[tb]
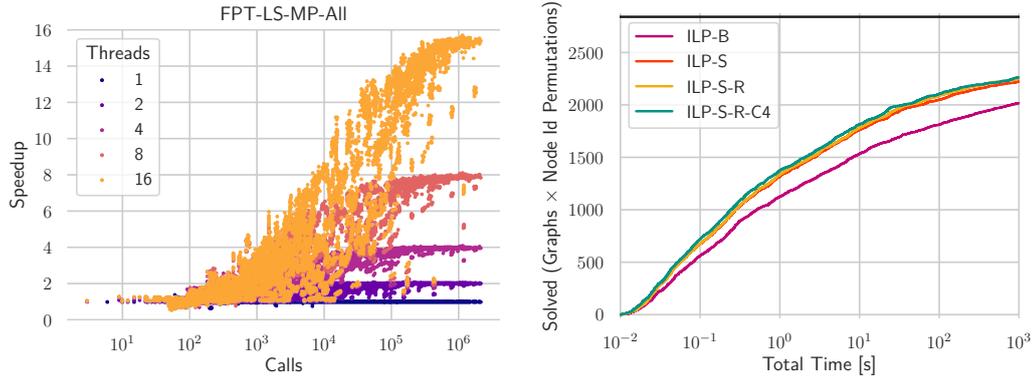

  \resizebox{0.5\textwidth}{!}{\input{fig/biospeedupmink20.pgf}}%
  \resizebox{0.5\textwidth}{!}{\input{fig/bio_gurobi_times_min_k_20.pgf}}%
  \caption{Speedup of \fpt\localsearch\bmost\all\mt{} and comparison of the different ILP variants on 16 (left) and 4 (right) node id permutations of the 716 COG graphs that require at least 20 edits.}\label{fig:cog-speedup-ilp}
\end{figure}

The left part of Figure~\ref{fig:cog-speedup-ilp} reports the speedup of \fpt\localsearch\bmostpruned\all\mt{} over its sequential counterpart \fpt\localsearch\bmostpruned\all{}, the sequentially fastest variant on the COG dataset.
We show the speedup with 1, 2, 4, 8 and 16 cores in comparison to the number of recursive calls and lower bound calculations.
For each graph and permutation, we plot the speedup on the last value of $k$ for which the sequential version of the algorithm terminated within the time limit.
With only few recursive calls we cannot expect a good speedup.
For a high number of recursive calls, \fpt\localsearch\bmostpruned\all\mt{} achieves almost perfect speedup for all numbers of cores on many graphs.
As the algorithm is executed with increasing values of $k$, for some graphs only the last value of $k$ needs a high number of calls and thus the overall speedup is not perfect even though in sum the number of calls is high.

\subsection{Variants of the ILP}\label{sec:exp:ilp}

\begin{figure}[tb]
  \resizebox{\textwidth}{!}{\input{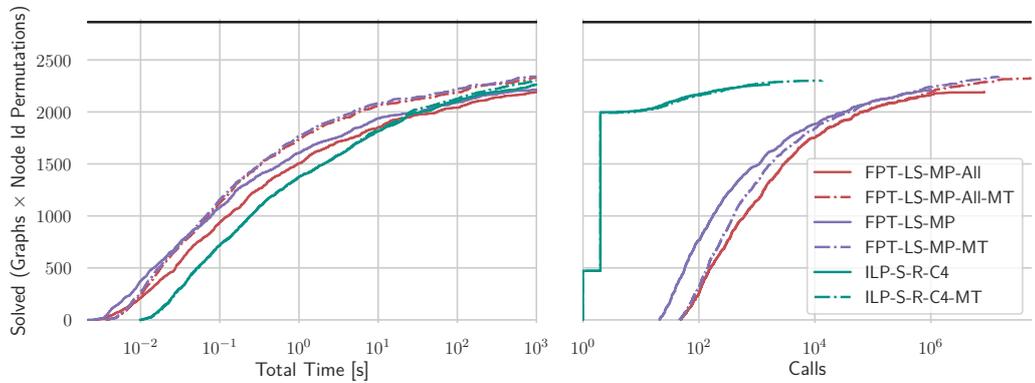}}%
	\caption{Comparison of the ILP to the FPT algorithm on 4 node id permutations of the 716 COG graphs that require at least 20 edits.}
	\label{fig:bio_final_comparison}
\end{figure}

Figure~\ref{fig:cog-speedup-ilp} (right) shows the impact of the different optimizations on the ILP, when enabled one after another.
We denote adding only one violated constraint by \single{}, adding constraints during relaxations by \relaxation, and specialized $C_4$ constraints by \cfour{}.
The baseline is \ilp\basic{}, where row generation always adds all violated constraints for intermediate solutions.

In the median, \ilp\single{} is just 5\% faster than \ilp\basic{}.
While on 95\% of the instances it is at most 20\% slower, it is more than 44 times faster on 5\% of them, which explains the gap in Figure~\ref{fig:cog-speedup-ilp}.
\ilp\single\relaxation{} is not faster in the median, but on 95\% of the instances at most 12\% slower and on 5\% it is at least 73\% faster.
The $C_4$ constraints make the ILP 12\% faster in the median, at most 26\% slower on 95\% and at least 95\% faster on 5\% of the instances.
With all optimizations, the ILP solves 568 graphs.
We also tried providing a heuristic solution from QTM~\cite{bhsw-fqte-15} to Gurobi, but the improvement was even smaller and disappeared in parallel.

Figure~\ref{fig:bio_final_comparison} compares the best ILP and FPT algorithms with and without \mt{} in terms of running time and recursive calls.
For the FPT algorithm, stopping at the first solution is not slower on 95\%, more than 52\% faster on 50\% and more than 3 times faster on 5\% of the instances.
Multi-threading incurs a measurable overhead.
Compared to \fpt\localsearch\bmostpruned\all{}, \fpt\localsearch\bmostpruned\all\mt{} is at most 16\% slower on 95\%, 78\% faster in the median and more than 12 times faster on 5\% of the instances.
When stopping at the first solution, this decreases to 24\% slower, 1\% faster and 10 times faster, as more branches that do not lead to a solution are explored in multi-threaded mode.
\fpt\localsearch\bmostpruned\mt{} is still 4\% faster than \fpt\localsearch\bmostpruned\all\mt{} in the median, at most 3\% slower on 95\% and at least 68\% faster on 5\% of the instances.

The parallel ILP is at most 5\% slower on 95\%, as fast in the median and more than 52\% faster on 5\% of the instances than the sequential ILP.
Thus, the parallelization helps the FPT algorithm more than the ILP.
A likely cause is that Gurobi needs much less search nodes than the FPT algorithm which offer less potential for parallelism -- on 50\% of the instances at least 185 times less, and on many graphs even just one or two, see Figure~\ref{fig:bio_final_comparison}.

The speedup of \fpt\localsearch\bmostpruned{} over \ilp\single\relaxation\cfour{} is at least 0.59 on 95\%, 3.25 in the median and at least 10.72 on 5\% of the instances.
For \fpt\localsearch\bmostpruned\all{}, this decreases to 0.29, 2.10 and 7.02.
In parallel, the speedups are 1.09, 3.41 and 16.45 for all solutions, and 1.34, 3.67 and 18.14 for the first solution.
Single-threaded, the ILP solves more instances within 1000 seconds than the FPT algorithm, indicating that for difficult instances better bounds are more important.
Overall, the FPT algorithm is often faster than the ILP, in particular in parallel and even when listing all solutions.

\subsection{Comparison to QTM}\label{sec:exp:qtm}

\begin{figure}[tb]
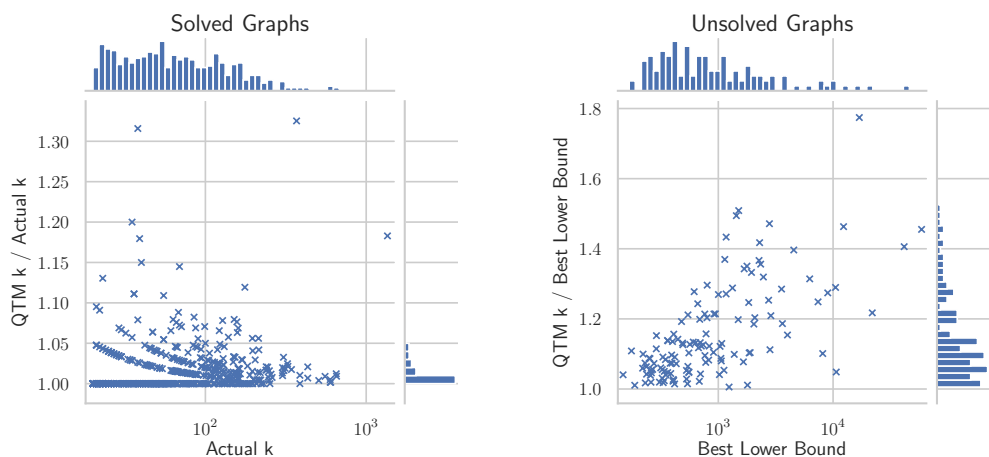

  \centering
  \resizebox{0.45\textwidth}{!}{\input{fig/bio_heur_qtm_solved.pgf}}%
  \qquad
  \resizebox{0.45\textwidth}{!}{\input{fig/bio_heur_qtm_unsolved.pgf}}%
  \caption{Comparison of heuristic solutions of QTM and the exact number of edits $k$ for solved graphs (left) or the best lower bound for unsolved graphs (right) achieved by the FPT algorithm or the ILP. For readability, we exclude one solved graph at $k = 64$, where QTM needed 202 edits.}\label{fig:cog-k}
\end{figure}

Figure~\ref{fig:cog-k} compares the results of the heuristic Quasi-Threshold Mover (QTM)~\cite{bhsw-fqte-15} with exact results for solved and the best lower bounds for unsolved graphs.
We use the maximum value of $k$ achieved for any permutation by \fpt\localsearch\bmostpruned\mt{} and by \ilp\single\relaxation\cfour{} with and without \mt{}.
If any of the algorithms solved the graph, we list it in the left part, otherwise in the right part.
For QTM, we report the minimum $k$ that QTM found over 16 runs.
Again, the plot excludes 3248 graphs that require less than 20 edits.
Of those, QTM solved 3172 exactly, 56 with offset 1, 15 with offset 2 and 5 with offset 3.
Of the remaining graphs, 588 are solved and 128 are unsolved.
Of the solved graphs, QTM solved 319 graphs exactly.
For none of the unsolved graphs, QTM matches the lower bound.
For 95\% of the 716 graphs, QTM needs at most 1.22 times the edits of the exact solution or the lower bound.

\subsection{Social Network Instances}\label{sec:exp:social}

\begin{table}[tbp]
	\caption{
		Overview of the social network graphs.
		Using the algorithms \fpt\localsearch\bmostpruned{} and \ilp\single\relaxation\cfour{} with 1 and 16 cores, we report the maximum $k$ that finished within 1000 seconds, and the minimum time over all permutations that is needed to find the first solution.
		In the case of \texttt{football}, we report the time needed to show that there is no solution with that $k$.}\label{tab:results_overview}
	\centering
	\begin{tabular}{lrrrrrrrrrr}
\toprule

           & && \multicolumn{4}{c}{FPT} & \multicolumn{4}{c}{ILP} \\
           & && \multicolumn{2}{c}{1 core} & \multicolumn{2}{c}{16 cores} & \multicolumn{2}{c}{1 core} & \multicolumn{2}{c}{16 cores} \\
     Graph &    n &    m &      k & Time [s] &        k & Time [s] &      k & Time [s] &        k & Time [s] \\
\midrule
    karate &   34 &   78 &     21 &     0.01 &       21 &     0.01 &     21 &     0.02 &       21 &     0.03 \\
    lesmis &   77 &  254 &     60 &     0.17 &       60 &     0.13 &     60 &     0.96 &       60 &     0.97 \\
 grass\_web &   75 &  113 &     34 &     1.81 &       34 &     0.21 &     34 &     2.91 &       34 &     2.83 \\
  dolphins &   62 &  159 &     70 &   126.54 &       70 &    18.57 &     70 &    23.81 &       70 &    12.10 \\
  \midrule
  football &  115 &  613 &    223 &   929.55 &      228 &   649.94 &    235 &  1000.01 &      237 &  1000.05 \\
\bottomrule
\end{tabular}

\end{table}

Table~\ref{tab:results_overview} shows an overview of the social networks with results for \fpt\localsearch\bmostpruned{} and \ilp\single\relaxation\cfour{}.
Both solve \texttt{karate} and \texttt{lesmis} in less than a second, and \texttt{grass\_web} within 3 seconds, with the FPT algorithm being faster.
Even though \texttt{lesmis} is both larger than \texttt{grass\_web} and requires 60 edits instead of 34, both algorithms are significantly slower on \texttt{grass\_web}.
This shows that their performance depends on the specific structure of the graph and not just the graph size and $k$.
For \texttt{dolphins}, the ILP is faster than the FPT algorithm.
For all graphs, the FPT algorithm scales better with the number of cores.
None of the algorithms can solve the \texttt{football} network.
We show that there is no solution for $k \leq 223$, $k \leq 228$ using the FPT algorithm with 1 or 16 cores respectively, and $k \leq 235$, $k \leq 237$ using the ILP with 1 or 16 cores respectively.
The previously best known upper bound was 251, computed with QTM~\cite{bhsw-fqte-15} in 2.5ms.
In 1000 seconds, the ILP shows a new upper bound of 250.
For the smallest three social networks, we verify that the best heuristic solutions in~\cite{bhsw-fqte-15} are exact.
QTM needs 72 edits on \texttt{dolphins}, whereas 70 edits are optimal.
Appendix~\ref{app:detailed-evaluation} contains a detailed analysis of the solution space with a focus on the community detection application.

\section{Conclusion}\label{sec:conclusion}

We have introduced optimizations for two different approaches to solving any $\mathcal{F}$-free edge editing problem.
We evaluate our optimizations for the special case of quasi-threshold editing on a set of 716 protein interaction graphs.
For the first approach, the FPT algorithm, we show that the combination of good lower bounds with careful selection of branches allows to reduce the running time by one to three orders of magnitude for 75\% of the instances.
For the second approach, an ILP, we evaluate several variants of row generation and show that they achieve small speedups.
We show that the FPT algorithm is slightly faster than the ILP, with a larger margin in parallel, and it can easily enumerate all optimal solutions.
For the heuristic editing algorithm QTM, we show that on 95\% of the instances, it needs at most 22\% more edits than our exact solutions or lower bounds indicate.

Comparing the structure of exact vs. heuristic solutions might give further insights how to improve heuristics.
Exact FPT algorithms could be further improved by better bounds, possibly based on LP relaxations.
As the COG benchmark set actually contains edit costs, an extension of our optimizations to the weighted editing problem could be investigated.
\bibliography{references_nocrossrefs}

\clearpage

\begin{appendix}

\section{Avoiding Redundancy}\label{app:redundancy}

Our algorithm maintains a global symmetric $n\times n$-bit matrix.
The entries in the matrix correspond to node pairs.
If a bit is set, the corresponding node pair must not be edited anymore.
We refer to these node pairs as \emph{blocked}.
In the following, we describe in detail how these optimizations that were introduced in~\cite{d-fpece-08} work.

\subsection{No Undo}\label{sec:no-undo}

Every solution that edits a node pair twice can be improved by not editing the node pair at all.
We exploit this observation by setting the bit corresponding to the performed edit when recursing.
When ascending from the recursion, we reset the bit.
The same optimization has also been used in~\cite{b-gzp-15} and is the basis of efficient branching rules for cluster editing~\cite{bbbt-gp-09}.

\subsection{No Redundancy}\label{sec:no-redundancy}

\begin{figure}
\begin{center}
\includegraphics{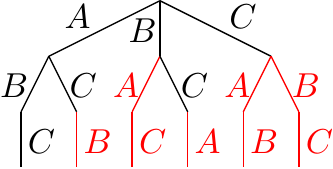}
\end{center}
\caption{Example Recursion Tree. Children are explored from left to right. The ``No Redundancy'' optimization prunes the red part.}
\label{fig:forbid-neighbor}
\end{figure}

A possible recursion tree is depicted in Figure~\ref{fig:forbid-neighbor}.
Note how multiple branches contain the same set of edits, but the edits appear in a different order.

We ensure that every branch enumerates a different set of edits, by unblocking a node pair only after all sibling edits have been explored.
In particular, this ensures that every solution is enumerated exactly once.
Consider the first recursion level of the example in Figure~\ref{fig:forbid-neighbor}.
The edits are explored in the following order: $A$, then $B$, and finally $C$.
Before descending into the branch of $A$, we set $A$'s bit.
After ascending from $A$'s branch and reverting the corresponding edit, the corresponding bit is not reset.
In addition to $A$'s bit, we set $B$'s bit and descend into $B$'s branch.
Finally, we ascend from $B$'s branch, leave its bit set, set $C$'s bit and descend into $C$'s branch.
After all edits in a recursion level are explored, all bits set in this level are reset, i.e., we reset $A$'s, $B$'s and $C$'s bit.

\section{Implementation Details}\label{app:implementation-details}

In this section, we document various details of our implementation.
We first describe our graph data structure and how we iterate over it.
In Section~\ref{app:implementation-details:subgraph-listing}, we describe how we list forbidden subgraphs.
We maintain subgraph counters that we describe in Section~\ref{app:implementation-details:subgraph-counters}.
In Section~\ref{app:implementation-details:lower-bound}, we describe for each lower bound algorithm how it is implemented using the aforementioned subgraph listing algorithms.
In Section~\ref{app:implementation-details:branching}, we describe the implementation of our branching strategy.
Finally, in Section~\ref{app:implementation-details:parallelization}, we describe our parallelization in detail.

We store our graph as an adjacency matrix with 1 bit per node pair.
To enumerate edges or neighbors, we use special CPU instructions to count leading zeros in a copy of a 64 bit block of this matrix.
We then remove the found 1-bit from the 64 bit block and count again.
If the current 64 bit block contains only zeros, we move to the next one.
All but the largest 24 graphs in our benchmark set have at most 320 nodes, thus requiring at most five 64 bit blocks per row of the matrix.
The largest graph has 8836 nodes and thus requires 139 64 bit blocks per row, but its average degree is also 64, therefore on average almost every second block contains a 1-bit.
Thus, for almost all of the graphs we consider, bit matrices seem an appropriate choice in terms of memory usage and enumeration efficiency.
For the larger graphs, adjacency arrays might be a better choice but as we are far from solving them, we did not further explore this.
Adjacency matrices have the advantage that we can easily combine multiple rows to list common neighbors or exclude neighbors of another node, a feature that we use for subgraph listing as described in the following.
Let $A[i,j]$ denote the entry of the adjacency matrix in row $i$ and column $j$.
We use matrix slice notation $A[:,j]$ to denote column $j$, and $A[i,:]$ to denote row $i$, i.e., the neighbors of node $i$.

\subsection{Subgraph Listing}\label{app:implementation-details:subgraph-listing}

To select subgraphs for branching and calculating lower bounds, we need to enumerate all forbidden subgraphs.
In preliminary experiments, we found that enumerating forbidden subgraphs on demand does not only require much less memory than storing them, but is also much faster.
Our implementation provides two methods for this: a global one that lists all forbidden subgraphs and a local one that lists all subgraphs containing a certain node pair.
The latter is required to efficiently implement our local search lower bound and the branching on most useful node pairs.
For simplicity, our descriptions focus on $\{C_4, P_4\}$-listing but our source code works for arbitrary $\{C_l, P_l\}$, $l \ge 4$ and $\{P_l\}$, $l \ge 2$.

\subparagraph{Global Listing.}
For the listing of all forbidden subgraphs, we enumerate all edges.
We consider each edge $\{u_2, u_3\}$ as the central edge and then enumerate edges $\{u_1, u_2\}$ in the outer loop, and $\{u_3, u_4\}$ in the inner loop, to complete the $P_4$ or $C_4$.
For listing candidates $u_1$, we directly exclude neighbors of $u_3$ by only iterating over $A[u_1,:] \land (\lnot A[u_3,:])$.
We list candidates $u_4$ analogously.
Fixing the central edge ensures each induced $P_4$ is listed exactly once.
We list each $C_4$ four times, which we use for trying different shareable node pairs for the packing lower bounds, as each edge deletion transforms the $C_4$ into a $P_4$.

\subparagraph{Local Listing.}
For the listing of forbidden subgraphs that contain a certain node pair $\{u, v\}$, we need to consider all positions of $\{u, v\}$ in the forbidden subgraph.
If $\{u, v\}$ is an edge, this means that apart from the case where $\{u, v\}$ is the central edge, we also need to consider the case where we extend the path twice on each side.
If $\{u, v\}$ is not an edge, the case where $\{u, v\}$ consists of the two degree-1-nodes of the $P_4$ can be omitted due to the optimizations discussed in Section~\ref{sec:cons-edit-last}.
We only need to find common neighbors $x \in A[u,:] \land A[v,:]$ of $u$ and $v$.
These are part of the central edge.
We try extending the path by one edge from $u$ and $v$ separately, i.e., iterate over $A[u,:] \land (\lnot A[x,:])$ and $A[v,:] \land (\lnot A[x,:])$.

\subparagraph{Listing For Lower Bounds.}
For lower bounds, we are only interested in forbidden subgraphs that do not contain any node pair that is already used in the lower bound.
We maintain a bit matrix $L$ where all node pairs that are already used in the bound are set to $1$.
By using $A[u,:] \land (\lnot L[u,:])$ instead of $A[u,:]$ for neighbors of $u$ and $(\lnot A[u,:]) \land (\lnot L[u,:])$ for non-neighbors, we can directly exclude these node pairs from listing.

\subparagraph{Excluding Specific Node Pairs From Listing.}
To branch on its node pairs or to check if a subgraph can be added to a lower bound, we need to enumerate its node pairs.
For this, we implicitly exclude blocked node pairs as well as $\{u_1, u_4\}$, which is the node pair of degree one in a $P_4$.
As mentioned before, we always list a $C_4$ four times, and thus omit a different edge $\{u_1, u_4\}$ in each enumeration.
This lets the lower bound algorithms select the best node pair to share or the branching strategy select the best node pair to exclude.

\subsection{Subgraph Counters}\label{app:implementation-details:subgraph-counters}

In our \texttt{Most} and \texttt{Most Pruned} branching strategies and the local search lower bound, we want to select the subgraph whose node pairs cover the most or least other forbidden subgraphs.
For this, we maintain a counter for each node pair in how many forbidden subgraphs it is contained.
Whenever a node pair is edited or blocked/unblocked we update the counters.
When blocking a node pair, we store its previous counter on a stack so that it can be easily restored when unblocking, and set the current counter to zero.
Note that our counters count a $C_4$ three times for edges and four times for non-edges due to listing the $C_4$ four times and omitting one of the edges each time.
We also maintain the sum of the subgraph counters to be able to quickly check if there are any forbidden subgraphs at all.

\subsection{Lower Bound Algorithms}\label{app:implementation-details:lower-bound}

Each of our lower bound algorithms has both a thread-state that is maintained once per thread and a call-state that is copied for every recursive call.
We compute an initial lower bound on the input graph and start our search for $k_{\text{opt}}$ from this bound instead of $0$.
The call-state is initialized once during this initial lower bound calculation and then used as initialization for all $k$s that we try.
For most algorithms, the call-state contains the previously calculated lower bound as an array of subgraphs (node tuples) in the packing.
We pass the call-state down into recursive calls, but not back up.
The rationale behind this is that we need to remove at most one forbidden subgraph from the bound when descending into recursion, whereas no longer blocked node pairs would force a lot of subgraphs to be removed when returning from a recursive call.

\subparagraph{Basic Bound.}
For the basic bound, we globally enumerate forbidden subgraphs $H$ and add $H$ to the packing $P$ if none of its node pairs are used by another graph in the packing.
This is done in each recursive call with an initially empty packing.
After the one pass, $P$ is inclusion-maximal.
We maintain a bit matrix $C$ for node pairs covered by the packing in the thread-state.
When adding $H$ to $P$, we mark its node pairs in $C$.
We also supply a reference of $C$ to the listing algorithm to skip subgraphs we cannot use.
As additional bits in $C$ are set during the listing, it does not skip all subgraphs we cannot use.
For example, the listing does not check the central node pair again.
In preliminary experiments this still gave a small speedup.

\subparagraph{Updates.}
For the basic bound with updates we pass the packing $P$ through the search tree.
Before descending into recursion, we remove the subgraph $H$ that contains the edited node pair $\{u,v\}$ from $P$, if it exists.
If possible, we add subgraphs $H'$ to $P$ that share a node pair with $H$ or contain $\{u,v\}$, to make $P$ inclusion-maximal.
This is done using the local listing.
Similar to before, the local listing skips some subgraphs touched by $C$.
We store a list of subgraphs in the packing in the call-state.
For memory efficiency, we do not store $C$ in the call-state but instead recompute it from scratch when modifying the bound.

\subparagraph{Local Search.}
The local search lower bound also just maintains the list of subgraphs used in the bound in its call-state.
The initial update works as described above.
To find candidates for replacing one subgraph by one or more subgraphs, we use the local listing.
In each round, we try to replace each subgraph $H$ in the packing once.
For this, we first remove $H$ from $P$ and then use the local listing on all node pairs of $H$ to obtain the set $R$ of subgraphs that could replace $H$.
From $R$, we obtain candidates that can be inserted together.
For each subgraph $H' \in R$ we first insert it into $P$, and then iterate over the rest of $R$, trying to insert.
If at least one additional candidate was found, we keep them in the packing.
Otherwise, we can only replace $H$ by $H'$.
With 70\% probability we take the $H'$ that covers the fewest other forbidden subgraphs, and with 30\% probability we choose a random one from $R$.

We apply several optimizations to speed up the search for additional candidates.
For each node pair of $H$, we store a separate list of candidates.
This allows us to skip candidates that use a node pair that is used by an already included candidate, without considering each candidate separately.
To avoid trying the same candidate twice, we also list candidates only for the first node pair they contain by excluding the already considered node pairs from the candidate search for subsequent node pairs.

\subparagraph{Min-Degree Heuristic.}
The min-degree heuristic is based on the independent set formulation where a subgraph is a node and two nodes are connected by an edge if the corresponding subgraphs share a node pair.
A good lower bound then corresponds to a large independent set.
For independent sets, the min-degree heuristic iteratively adds the node with the smallest remaining degree to the independent set and then deletes it and its neighbors from the graph.
Instead of explicitly constructing this graph model, we translate this formulation back to forbidden subgraphs.

We iteratively add subgraphs to the bound whose node pairs are shared with the least number of subgraphs that can still be added to the bound.
To implement it, we need to explicitly maintain the ``degree'' of every subgraph in a priority queue as it is changing over time as more and more subgraphs are added to the bound.
For this, we (temporarily) store an explicit list of all forbidden subgraphs that we obtain through global listing.
Further, for every node pair we store a list of subgraphs it is part of by storing their indices in the list of subgraphs.
Similarly, we store these list indices in the priority queue.
This allows to efficiently identify the elements that need to be updated or removed from the priority queue.
For a subgraph $H$, we initially use the sum over all node pairs of $H$ of the number of subgraphs that contain the node pair as key.
This might count the same subgraph several times, but is more efficient to calculate.
Preliminary experiments showed that this is faster than calculating the actual number of subgraphs with whom $H$ shares a node pair.

Whenever we take a subgraph $H$ from the priority queue and add it to the bound, we need to remove its neighbors from the priority queue and update their neighbors' degrees accordingly.
To obtain the neighbors $N$ of $H$, we iterate over $H$'s node pairs and list all subgraphs they are part of.
We remove each subgraph $n \in N$ from the priority queue and for each of $n$'s neighbors, we decrement its key in the priority queue by one.

For the priority queue, we use a bucket priority queue.
As nodes only need to be moved between adjacent buckets, we can maintain all buckets in one large array and move elements between buckets by swapping them to the boundary and then adjusting the boundary.

\subparagraph{LP Relaxation.}
The LP for our \emph{LP bound} corresponds exactly to the ILP formulation shown in Section~\ref{sec:ilp} with the optimization of omitting one node pair as described in Section~\ref{sec:ilp:specialc4}.
Our main goal for the implementation of the LP bound was to have a comparison with a lower bound algorithm that is guaranteed to prune at least as good as the packing-based lower bounds.
For this reason, we ensure that the LP always contains the constraints that correspond to forbidden subgraphs that could also be used in the packing-based lower bound.
Similar to the ILP, we initialize the LP with all constraints that correspond to forbidden subgraphs in the input graph.
Whenever we edit or block a node pair, we fix the value of its corresponding value to 1 or 0, depending on whether it is now connected by an edge or not.
When a node pair is edited, it is always blocked and thus all constraints that correspond to forbidden subgraphs that no longer exist in the edited graph are trivially fulfilled and there is thus no need to remove them explicitly.
After each edit, we add constraints that correspond to forbidden subgraphs that contain the edited node pair.
After undoing an edit, we remove them again, as the LP solver slows down when the LP contains a lot of constraints.

\subsection{Branching Strategies}\label{app:implementation-details:branching}

As described in Section~\ref{sec:most}, we want to prefer subgraphs that contain at most one non-blocked node pair, as we know this edit has to be applied.
In a call-state like those of the lower bound algorithms, we store both this list of subgraphs, and a flag indicating whether the branch can be pruned.
Such subgraph can only appear when blocking or editing a node pair $\{u, v\}$.
Every time this happens, we enumerate all subgraphs $H$ containing $\{u, v\}$.
If $H$ contains exactly one non-blocked node pair, we store it.
If $H$ contains only blocked node pairs, the current branch can be pruned immediately because $H$ cannot be destroyed.
Before the branching strategy selects a forbidden subgraph, we first check if the flag is set and return an empty list of node pairs, if so.
Otherwise, we iterate over the list of subgraphs in the call-state and return the only non-blocked node pair of the first subgraph of the list where this node pair has not been edited yet.
Note that two of these subgraphs might contain the same non-blocked node pair, thus after the first of them has been selected, the second becomes invalid.

Only if the flag is not set and there is no subgraph with exactly one non-blocked node pair, we apply the actual branching strategy.
In our \texttt{Most} and \texttt{Most Pruned} branching strategies, we avoid listing all forbidden subgraphs.
Instead, we first identify those node pairs that are part of the maximum number of subgraphs using the subgraph counters introduced in Section~\ref{app:implementation-details:subgraph-counters}.
Due to our lexicographical ordering, we are only interested in those subgraphs that contain these node pairs.
We use our local listing to enumerate them and select the maximum as described in Section~\ref{sec:most}.
The output of the branching strategy is a sorted list of node pairs and a flag whether the graph is solved.
We set this flag if the sum of the subgraph counters is zero.

\subsection{Parallelization}\label{app:implementation-details:parallelization}

In our parallelization, different threads explore different branches of the search tree.
We maintain a global queue of work packages that represent roots of unexplored branches.
To achieve a scalable parallelization we want to generate few work packages that have a lot of recursive calls left.
Due to our optimizations, we cannot know in advance how many calls are left for a certain branch.
Even branches that start at the root of our recursion tree might be pruned after a single call.
Therefore, we need to generate work packages as we explore the search tree, i.e., employ work-stealing.

Each work package contains the number of remaining edits, the graph, the blocked node pairs, the subgraph counters and the call-states of the lower bound and the branching strategy (can be empty).
Hence, creating a work package for every call is too expensive, as we would need to copy these data structures, whose memory consumption is quadratic in the number of nodes.
Passing them through the search tree, and updating on-the-fly is fast, but a work package constitutes the root of a new search tree and thus requires a copy.
Therefore, we only create work packages when the global work queue contains less work packages than the number of threads.

When a worker finishes one recursion tree, it takes another work package from the queue.
If there is none left, it waits until either work becomes available or the algorithm is finished.
The latter is indicated either by the fact that no thread has a work package anymore (we keep a counter how many threads are currently working), or a global flag that is set when the first solution has been found and not all solutions shall be listed.
This flag is also checked in each recursive call to ensure that if one thread finds a solution, all other threads terminate.

At the beginning of every recursive call, we check if work packages shall be generated.
A simple approach would be to split the recursive calls of the current search tree node into work packages.
Unfortunately, this does not scale well, as we would predominantly create work packages on deeper recursion levels where only few edits remain.
Instead, we split off unexplored branches from the top of the current recursion tree, where we hope the most work is left.
For this, we explicitly maintain the current recursion path of each worker thread.

Each element of the path contains the node pairs to branch on, the call-states of the bound and the branching strategy for each of these branches and an index that indicates the next branch to be explored.
After potentially generating work packages, we invoke the lower bound calculation.
We then check if the recursive call can be pruned because of the bound, because there are no more edits left or because a solution has been found.
If not, we create its element in the path.
For this, we obtain the node pairs for the next recursion level from the branching strategy.
We then create copies of the call-states for all of them and update them such that each of them can be directly used to create a work package.
This ensures that work package generation, which happens inside a global lock, is quick and does not need to update call-states.
For the early pruning (see Section~\ref{sec:prune}), we also directly check if calls can be pruned and if yes, we directly remove them from the node pairs.

For the actual recursive calls, we iterate over the node pairs in the element of the path.
We advance the index that indicates the next call and execute an actual recursive call with the call-state for that node pair.
After all recursive calls finished, we remove the element from the path.
It is possible that during a recursive call on a lower level work packages have been generated for the remaining node pairs.
In this case, the path will be empty when the recursive call returns.
We check for this, and then return directly instead of continuing with the remaining node pairs.

For our recursive calls, we also update a copy of the graph, the blocked node pairs and the subgraph counters that are used for calculating lower bounds and the branching strategy.
Additionally to this copy that represents the state at the \emph{bottom} of our path, we also maintain a copy that corresponds to the state at the \emph{top} of the path that is used for generating work packages.

To generate work packages, we first advance the \emph{top} state to the next node pair that has not been used for a recursive call.
For all remaining node pairs of the top element of the path we generate a separate work package using the top state and the call-state that is stored in the path's element.
Then, we remove the top of the path.
This continues with the new top of the path, until either the recursion path is empty or a sufficient number of work packages ($2x$ number of threads) are in the queue.

Note that we generate work packages before creating the element in the path that corresponds to the current recursive call.
Hence, the generating thread still has work left, even if the recursion path becomes empty.
This is to avoid that a thread immediately needs to get another work package after putting work into the global queue.

\section{Evaluation of the Found Solutions}\label{app:detailed-evaluation}

The FPT algorithm offers the possibility to list all solutions exactly once.
In this section, we show the number of solutions found for the COG dataset.
Further, we examine the solutions that are found on the four solved social network instances in detail, comparing them concerning the community detection application of quasi-threshold editing.

\begin{figure}[tb]

  \resizebox{\textwidth}{!}{\input{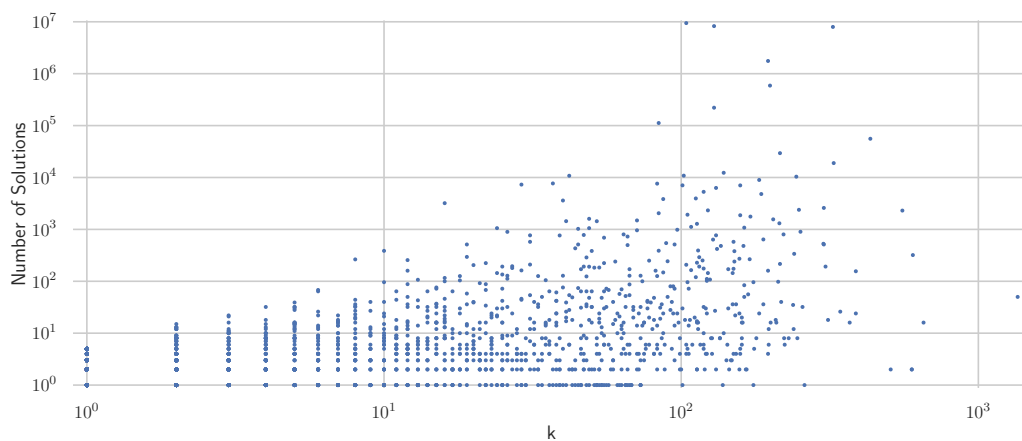}}
	\caption{Number of solutions for all solved graphs of the COG dataset sorted by $k$ with $k > 0$.}\label{fig:cog-solutions}
\end{figure}

Figure~\ref{fig:cog-solutions} shows the number of found solutions for all solved graphs of the COG dataset.
We plot the number of solutions for the graphs grouped by $k$ to see if there is any correlation between $k$ and the number of solutions.
For some graphs with $k > 100$, there are over a million solutions found.
However, there seems to be no strong correlation between $k$ and the number of solutions found.
Nevertheless, this shows that in many cases there is not one clear solution for a graph.

\begin{figure}[tb]
  \hfill
  \includegraphics[width=0.4\textwidth]{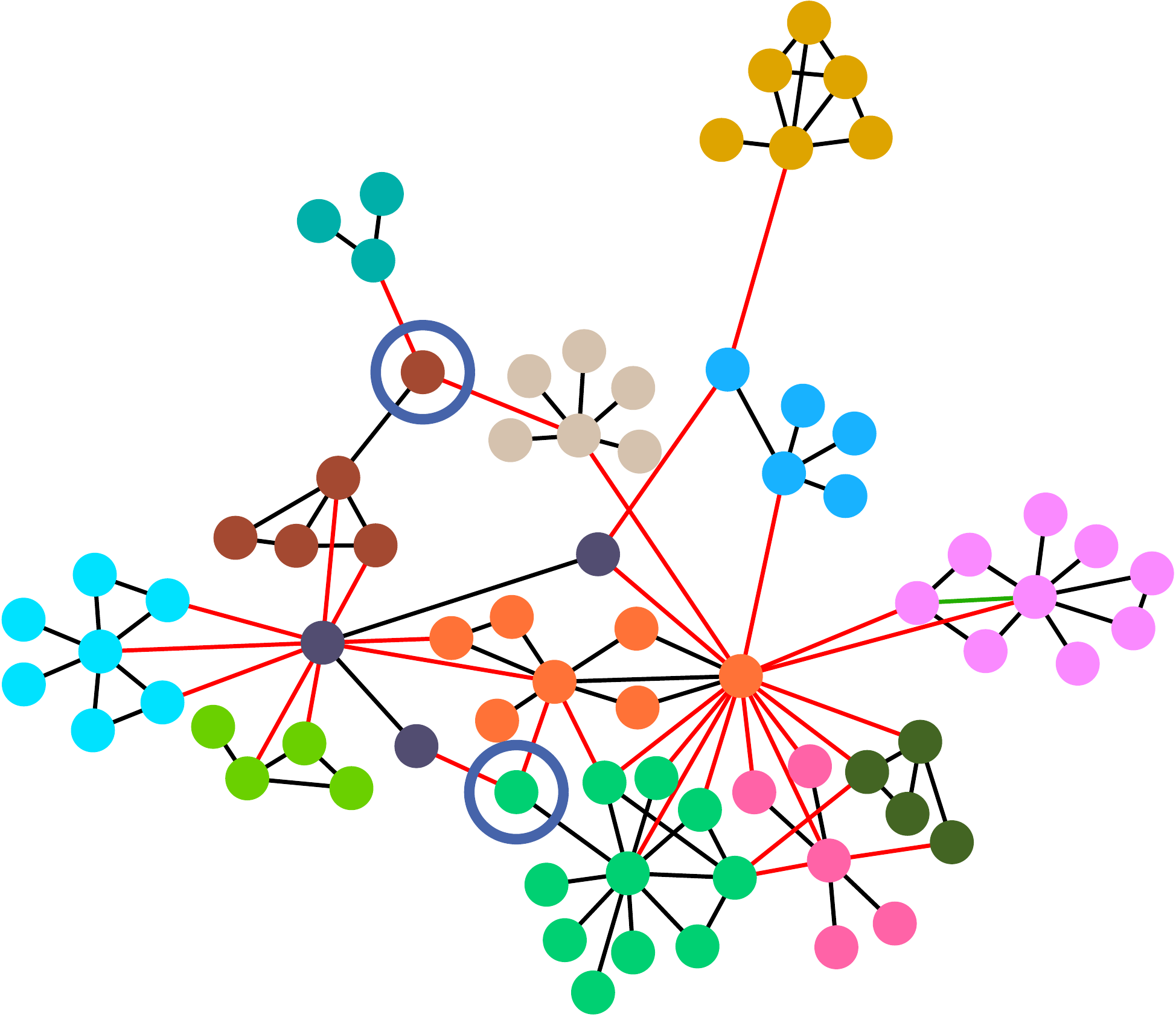}
  \hfill
  \includegraphics[width=0.4\textwidth]{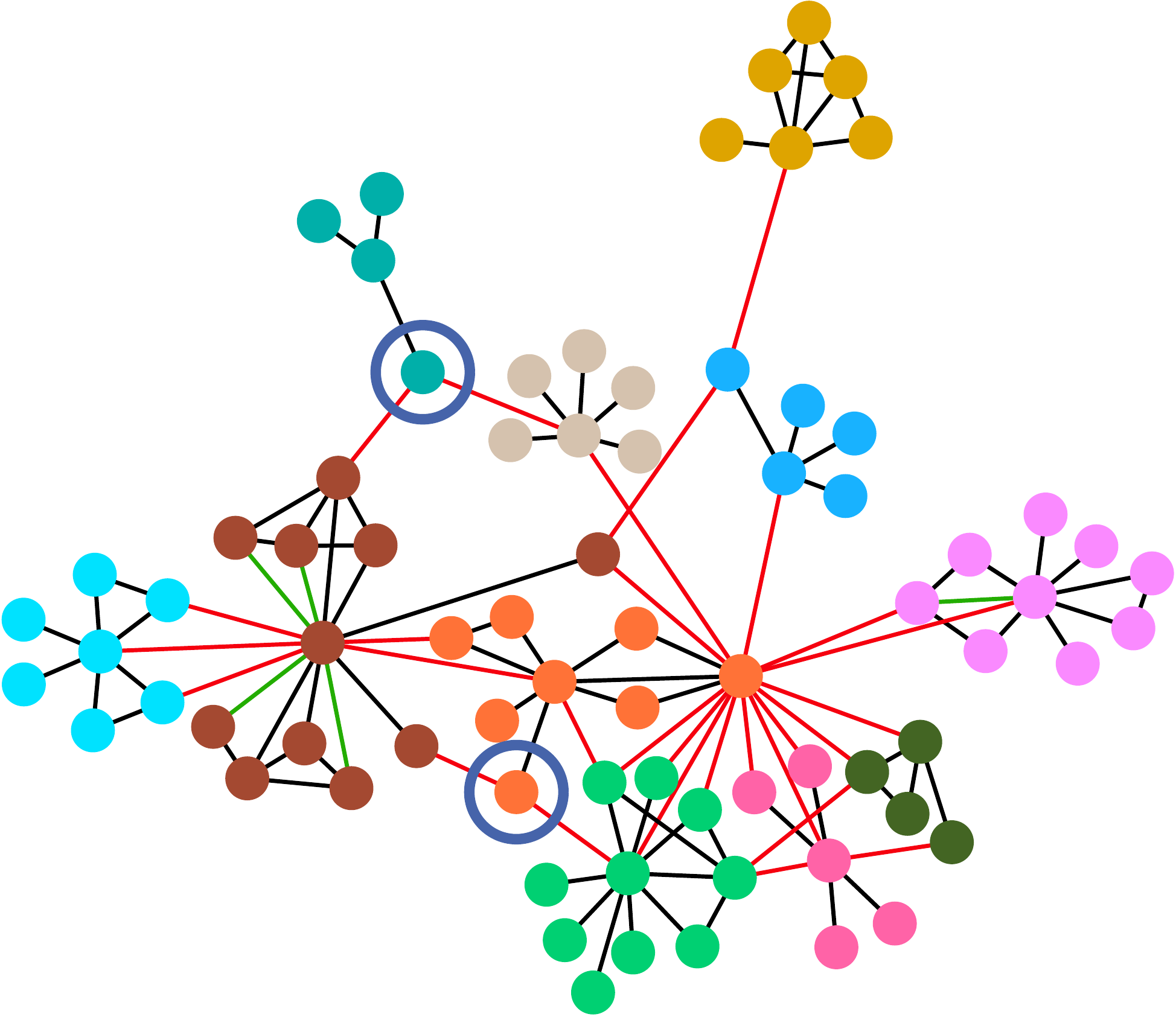}
  \hfill
  \caption{Two solutions of \texttt{grass\_web}. Red edges have been deleted, green edges have been inserted. Nodes are colored by connected component in the edited graph. The two blue circles denote two nodes that changed clusters.}\label{fig:grass_web_solutions}
\end{figure}

To examine how this affects the graph clustering application suggested by~\cite{ng-f-13}, we take a closer look at the solutions for the four solved social networks using NetworKit~\cite{ssm-nka-16}.
Table~\ref{tab:solutions} summarizes the found solutions.
The number of found solutions ranges from 24 on \texttt{dolphins} up to 3006 for \texttt{grass\_web}.
Nastos and Gao~\cite{ng-f-13} propose to use the connected components of the edited graphs as clusters.
While two closest quasi-threshold graphs might use different edits, they can still induce the same clustering as some edits only affect edges inside clusters.
Therefore, we also examine the number of different clusterings found.
For \texttt{karate}, the 896 solutions induce only 12 different clusterings, while for \texttt{grass\_web}, there are 2250 different clusterings induced by 3006 solutions.
This shows that there can be quite some variance in terms of the found clusterings even between exact solutions.
The number of clusters remains rather stable, on the other hand.
Figure~\ref{fig:grass_web_solutions} shows two solutions of \texttt{grass\_web} with two different clusterings.
The brown cluster in the right solution is split into the brown, bright-green and dark-blue clusters in the left solution.
Further, two nodes that are marked by a blue circle in the figures switch their cluster assignment.

\begin{table}[tb]
  \caption{Summary of the solutions found. For each graph, we report the number of different solutions, the number of different induced clusterings, the minimum and maximum number of clusters in the different solutions, the number of insertions and deletions common to all solutions, the number of clusters obtained when just applying the common edits, the total number of different insertions and deletions and the number of clusters obtained when intersecting all found clusterings.}\label{tab:solutions}
  \centering
\begin{tabular}{lrrrrrrrrrrr}
\toprule
   &        & \#Solu- &  \#Clus- & \multicolumn{2}{c}{\#Clusters} & \multicolumn{3}{c}{Common} & \multicolumn{2}{c}{Union} \\
     Graph &   k &  tions & terings &       Min & Max &   Ins. & Del. & Clus. &  Ins. & Del. & Clust. \\
\midrule
    karate &  21 &    896 &      12 &         2 &   4 &      0 &   11 &     2 &    13 &   27 &     7 \\
 grass\_web &  34 &   3006 &    2250 &        11 &  14 &      1 &   11 &     2 &    11 &   45 &    22 \\
    lesmis &  60 &    384 &     192 &         8 &  12 &      4 &   45 &     6 &    10 &   63 &    16 \\
  dolphins &  70 &     24 &       8 &        12 &  13 &      5 &   56 &     9 &    11 &   71 &    16 \\
\bottomrule
\end{tabular}
\end{table}

To see if there is something all solutions can agree on, we examine the edits that are common to all solutions.
Out of the 21 edits necessary for \texttt{karate}, there are 11 common edge deletions.
These induce a cut into two clusters, which is also the cut found in~\cite{z-ifmcf-77}.
For \texttt{grass\_web}, there are 1 edge insertion and 11 edge deletions common to all solutions which also only split the graph into two parts -- compared with the necessary 34 edits and the up to 14 clusters in each solution.
For \texttt{lesmis}, there are 4 edge insertions and 45 edge deletions common to all solution which induce 6 clusters -- this shows a structure that is a lot more stable.
On \texttt{dolphins}, there are 5 edge insertions and 56 edge deletions common to all solutions, i.e., each solution only adds 9 further edits.
These common edits already induce 9 clusters which is close to the 12 or 13 clusters found in the individual solutions.

Additionally, we look at the number of edits in the union of solutions.
For all graphs there are more edge deletions than insertions.
Even if all edge insertions were in a single solution, in all graphs but \texttt{karate} there were more than two times more edge deletions than insertions.
Further, we calculate the intersection of all found clusterings to obtain the largest clusters that are not split in any solution.
For \texttt{karate}, this gives us 7 clusters that split both of the two parts into further parts.
For \texttt{grass\_web}, we even obtain 22 clusters (compared with at maximum 14 clusters in an individual solution).
For \texttt{lesmis} and \texttt{dolphins}, we obtain 16 clusters, i.e., a value relatively close to the up to 12 or 13 clusters that are found in individual solutions.

This analysis shows the power of being able to enumerate all solutions.
We can not only determine how stable the clustering structure of a graph is, we can also obtain smallest components on which all different solutions agree -- or large clusters, where all solutions agree that they should be split.
This could also be used to obtain overlapping clusters by assigning nodes that frequently change between clusters to several clusters.

\end{appendix}

\end{document}